\documentclass[a4paper]{article}
\usepackage{Odyssey2024}
\usepackage{epsfig,amssymb,amsmath}
\ninept
\usepackage{multirow}
\usepackage{pifont} 
\usepackage{amsmath}
\usepackage{svg}
\usepackage{graphicx}
\usepackage{titlesec}
\usepackage{amsfonts}
\usepackage{amssymb}
\usepackage{hyperref}
\usepackage{caption}
\usepackage{array}
\usepackage{booktabs} %

\titleformat{name=\section,numberless}[block]{\normalfont\Large\bfseries\centering}{}{0pt}{}

\usepackage{sectsty} 
\DeclareMathOperator*{\argmin}{arg\,min}

\title{Leveraging Speaker Embeddings in End-to-End Neural Diarization for Two-Speaker Scenarios}

\name{Juan Ignacio Alvarez-Trejos, Beltrán Labrador, Alicia Lozano-Diez}

\address{AUDIAS - Audio, Data Intelligence and Speech  \\
Universidad Autónoma de Madrid \\
{\small \tt \{juani.alvarez, beltran.labrador, alicia.lozano\}@uam.es} }

\begin{document}
\maketitle

\begin{abstract}
End-to-end neural speaker diarization systems are able to address the speaker diarization task while effectively handling speech overlap. This work explores the incorporation of speaker information embeddings into the end-to-end systems to enhance the speaker discriminative capabilities, while maintaining their overlap handling strengths. To achieve this, we propose several methods for incorporating these embeddings along the acoustic features. Furthermore, we delve into an analysis of the correct handling of silence frames, the window length for extracting speaker embeddings and the transformer encoder size. The effectiveness of our proposed approach is thoroughly evaluated on the CallHome dataset for the two-speaker diarization task, with results that demonstrate a significant reduction in diarization error rates achieving a relative improvement of a 10.78\% compared to the baseline end-to-end model.

\end{abstract}

\section{Introduction}
\label{sec:intro}
Speaker diarization, a crucial task in multiple speakers scenarios, involves identifying different speaker segments within audio recordings \cite{Park2022}. Traditional approaches typically segment audio into speech and non-speech chunks using Voice Activity Detection (VAD) \cite{8461921} before extracting speaker features from each speech segment. Utilizing VAD is essential as speaker embedding extractors are not typically trained to effectively represent silence. Among the commonly used speaker representations are i-vectors \cite{Ibrahim2018, 7078610}, d-vectors \cite{10.1109}, x-vectors \cite{7953094} and ECAPA-TDNN x-vectors \cite{Dawalatabad2021ECAPATDNNEF}, which encapsulate information about speaker identity, speech style, and other related attributes \cite{Raj2019ProbingTI}.

In the traditional modular diarization systems, these speaker embeddings are then clustered using conventional clustering algorithm such as K-means \cite{dimitriadis17_interspeech}, Agglomerative Hierarchical Clustering \cite{7078610, 7953094}, spectral clustering \cite{8462628} and Bayesian HMM clustering of x-vector sequences (VBx) \cite{landini2022bayesian}. However, traditional diarization strategies often struggle with overlapping speech, where multiple speakers talking simultaneously pose a challenge to isolate and cluster individual speaker information. Also, these speaker representations are often extracted using a fixed-size sliding window. The window size can significantly affect the diarization performance: while using larger window sizes helps capture more speaker information, it increases the likelihood of including multiple speakers, complicating clustering approaches. In contrast, smaller window sizes reduce this risk, but may not provide enough data to extract reliable speaker information. This trade-off between capturing speaker-specific details and minimizing overlap-induced errors underscores the importance of carefully selecting window sizes in diarization systems \cite{9747450}. 

To address the challenge of speaker overlap, recent advancements in speaker diarization have introduced end-to-end models, treating the task as a multi-label classification problem \cite{Fujita2020EndtoEndND, SERAFINI2023101534}. Models like the self-attentive end-to-end neural diarization (SA-EEND) \cite{Fujita2019EndtoEndNS, 9003959} integrate various aspects like voice activity and overlap detection, showing promising results at handling these issues natively. However, these models encounter difficulties with scenarios involving varying numbers of speakers. To overcome this limitation, newer models like end-to-end neural diarization with encoder-decoder attractors (EEND-EDA) \cite{Horiguchi2020EndtoEndSD, eendeda} have been developed to adapt dynamically to a variable speaker number. Additionally, hybrid systems combining end-to-end approaches with clustering algorithms have emerged achieving improved performance across diverse scenarios \cite{9414333, 10095589, Chen2023, 10022924, 10095370}. Similarly, in \cite{xia2022turn,10094955} speaker turns are detected at word level by an automatic speech recognition (ASR) system to subsequently perform speaker clustering diarization. 

Another method to tackle the diarization task is to condition voice activity detection with speaker embedding information.  This approach leverages speaker embeddings like i-vectors or x-vectors, concatenating them with acoustic features of speech segments and feeding them into the model.  Systems such as Target-Speaker Voice Activity Detection (TS-VAD) \cite{Medennikov2020} utilize this technique. Additionally, methods described in \cite{9747772, 10095185} demonstrate the effectiveness of jointly training EEND and TS-VAD components.

In this study, inspired by the capabilities of TS-VAD, we aim to enhance the speaker discriminative potential of the EEND-EDA diarization system by incorporating speaker information extracted from a pre-trained ECAPA-TDNN \cite{Dawalatabad2021ECAPATDNNEF} x-vector extractor. 

We have explored different ways to integrate the speaker embeddings at various points of the architecture, aiming to help the model to better leverage the speaker information. Additionally, we have focused on critical speaker embedding extraction hyperparameters such as the window size, for optimal performance in this task. Furthermore, to address the potential limitations of speaker embeddings in representing silent segments, we applied an oracle VAD to the speaker embeddings during the diarization training process, ensuring a better handling of silent segments within the model. Finally, we used an external VAD during testing for a comprehensive evaluation independent of oracle VAD labels.

The rest of this paper is structured as follows: Section \ref{sec:exps} explains the speaker diarization pipeline, detailing the architecture and training scheme of the proposed systems. Section \ref{sec:setup} presents the experiments conducted and their corresponding results are shown in Section \ref{sec:results}. Finally, the paper concludes with a summary in Section \ref{sec:conc}.

\section{Methods}
\label{sec:exps}

\subsection{Review of End-to-End Neural Diarization with Encoder-Decoder Attractors}
\label{sec:eend}
Here we provide a concise overview of the end-to-end diarization framework known as EEND-EDA \cite{Horiguchi2020EndtoEndSD}. The EEND module processes input data comprising a sequence of log-scaled Mel-Filterbank features $X = [x_1, x_2, ..., x_T] \in \mathbb{R}^{T \times F}$, where $T$ is the sequence length and $F$ the feature dimension. These features are passed through either bi-directional long short-term memory (BLSTM) \cite{Fujita2020EndtoEndND}, Transformer \cite{Fujita2019EndtoEndNS, 9003959}, or Conformer \cite{conformer} encoders to obtain embeddings $e = [e_1, e_2, ..., e_T] \in \mathbb{R}^{T \times D}$ at each time step.

To determine a flexible number of attractor points from variable-length embedding sequences, an LSTM-based encoder-decoder architecture is employed. The embedding sequence $(e_t)^{T}_{t=1}$, with $D$ dimensions, is fed into the unidirectional LSTM encoder. 

\begin{equation}
    h_0, c_0 = LSTM_{encoder}(e_1, ..., e_T)
\end{equation}

Attractors are generated during the decoding process.

\begin{equation}
    h_s, c_s, a_s = LSTM_{decoder}(h_{s-1}, c_{s-1}, 0)
\end{equation}

These attractors are defined as $A=[a_1, ..., a_s] \in \mathbb{R}^{D \times S}$ where $S$ is the number of speakers. The probability of the presence of an attractor $a$ is determined using a fully-connected layer with a sigmoid activation function, defined as:

\begin{equation}
    p_s = \dfrac{1}{1+exp(-(w^Ta_s+b))}
\end{equation}

Here, $w$ and $b$ represent the trainable weights and bias of the fully-connected layer, respectively. During training, ground truth labels are defined based on the actual number of speakers $S$.

\begin{equation}
l_s = \begin{cases}
1 & \text{if } s \in \{1, \ldots, S\} \\
0 & \text{if } s = S + 1
\end{cases}
\end{equation}

The attractor existence loss $L_\alpha$ is calculated between $\hat{y}$ and the ground truth labels $y$ using the binary cross entropy defined as:

\begin{equation}
    H(y_t, \hat{y}_t) := \sum_{s} \left( -y_{t,s} \log \hat{y}_{t,s} - (1 - y_{t,s}) \log (1 - \hat{y}_{t,s}) \right)
\end{equation}

Therefore:

\begin{equation}
    L_\alpha = \dfrac{1}{1+S}H(l,p)
\end{equation}

Finally, the diarization loss is computed using the Permutation Invariant Training (PIT) scheme \cite{7952154}. This loss is calculated between $\hat{y}_t$ and the ground truth labels $y_t = [y_{t,1} ..., x_{t,S}]^T \in \{0,1\}^S$ as:

\begin{equation}
L_d = \frac{1}{TS} \argmin_{\phi \in \text{perm}(1,...,S)} \sum_{t=1}^{T} H(y_{t}^{\phi}, \hat{y}_t)
\end{equation}

The final loss is defined by the diarization loss and the attractor existence loss.

\begin{equation}
    L = L_d + \alpha L_\alpha
\end{equation}

Where $\alpha$ is the weighting parameter for the attractor loss and it has a value of $1$ for training and $0.1$ for adaptation.

\subsection{ECAPA-TDNN speaker embedding extractor}

The ECAPA-TDNN architecture \cite{Dawalatabad2021ECAPATDNNEF} builds upon the well-established x-vector topology \cite{7953094} and integrates several enhancements to produce more robust speaker embeddings. During training, the network is optimized to directly minimize the cosine distance between speaker embeddings by optimizing the AAM-
softmax loss \cite{8953658}. This approach makes it less dependent on complex scoring backends such as Probabilistic Linear Discriminant Analysis (PLDA) \cite{ioffe2006probabilistic}. The ECAPA-TDNN model is trained using data from VoxCeleb1 and VoxCeleb2 \cite{nagrani2017voxceleb,chung2018voxceleb2}, with additional data augmentation from the RIRs~\footnote{https://www.openslr.org/28/} and MUSAN~\cite{Snyder2015} datasets, further enhancing its performance and generalization capabilities.

\begin{figure}[t]

\includegraphics[width=\columnwidth]{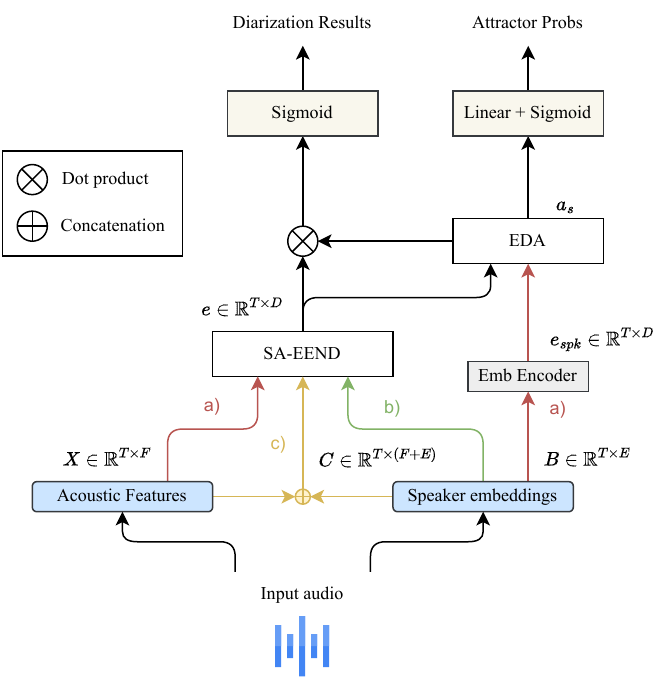}
\caption{{\it Architecture of the EEND-EDA model with our different proposed methods to integrate speaker embeddings into the system: a) Speaker embeddings into EDA module described in section \ref{method_a}, b) Speaker embeddings into SA-EEND encoder described in section \ref{method_b}, and c) Concatenation of speaker embeddings and MFbank into SA-EEND encoder as described in section \ref{method_c}.}}
\label{spprod}
\end{figure}

To ensure alignment between acoustic features and speaker embeddings, we construct two sequences with the same number of frames.  The acoustic feature sequence is $X = [x_1, x_2, ..., x_T] \in \mathbb{R}^{T \times F}$, and the corresponding speaker embedding sequence is $B = [b_1, b_2, ..., b_T] \in \mathbb{R}^{T \times E}$. We use the same acoustic features as in~\cite{Horiguchi2020EndtoEndSD}, and we adjust the sliding window of the ECAPA-TDNN embedding extractor to match the number of frames of the acoustic feature sequence.

\subsection{Proposed methods}

We propose three methods to combine the acoustic features $X$ and the speaker embedding $B$ sequences as input to the SA-EEND-EDA model. 
In Figure \ref{spprod}, the three proposed methods are depicted, distinguishable by the color of the arrows (i.e., red arrows indicate method a), where Mel-Filterbank (MFbank) goes to the SA-EEND module and the speaker embeddings to the EDA module).

\subsubsection{Speaker embeddings into EDA module}
\label{method_a}
As previously discussed in Section \ref{sec:eend}, the original encoder-decoder attractor (EDA) module receives as input the output vector sequence from the last SA-EEND transformer encoder block. This vector sequence, is then used for both the diarization task (optimized using the PIT loss) and the speaker identification or existence task within the EDA module. Specifically, the EDA aims to identify the active speakers within the corresponding frame segment. Once the attractors are calculated, they are used to condition the computation of the diarization posteriors applying the dot-product operation with the same encoder output vector sequence $e$. Our objective is to examine the behavior of the model by directly introducing the speaker embeddings $B$ calculated with the ECAPA-TDNN extractor into the EDA module as shown in Figure \ref{spprod} a). These embedding sequence $B$ are encoded with a single transformer encoder block to maintain the original input dimension of the EDA module, ensuring that:

\begin{equation}
    e_{spk} = \textit{Emb Encoder}(B) \in \mathbb{R}^{T \times D}
\end{equation}

In this setup, the SA-EEND transformer encoder takes just the Mel-Filterbank acoustic feature sequence $X$ as input. Thus, we aim to determine if including the speaker embedding information in the EDA module helps to better determine the attractors and represent speaker variability.

\subsubsection{Speaker embeddings into SA-EEND encoder}
\label{method_b}

We also explore how effective speaker embeddings $B$ are on their own as input to the end-to-end speaker diarization system, without using any acoustic features, and if the model is able to refine the speaker discriminative information within these embeddings for the task of diarization. To do this, we use these ECAPA-TDNN embeddings $B$ as input features, extracted at the same time resolution as $X$. In this setup, as in the SA-EEND-EDA baseline \cite{Horiguchi2020EndtoEndSD}, encoder output vector sequence $e$ are directly fed into the EDA module as shown in Figure \ref{spprod} b).

\subsubsection{Concatenation of speaker embeddings and MFbank into SA-EEND encoder}
\label{method_c}

Taking inspiration from the TS-VAD model's approach of combining speaker embeddings with acoustic features, our setup aims to work on this idea by concatenating the ECAPA-TDNN embedding sequence $B$ with Mel-filterbank acoustic feature sequence $X$, and using it as input to the EEND-EDA model. This concatenated sequence is denoted as:

\begin{equation}
C = X \oplus B \in \mathbb{R}^{T \times (F+E)}
\end{equation}

Where $\oplus$ refers to the concatenation operation. With this configuration, the model can utilize both feature sequences simultaneously, leading to an integration of acoustic and speaker-related information. This setup is illustrated in Figure \ref{spprod} c).

\subsection{Silence frames handling}

Speaker embeddings are not typically trained to effectively discriminate silence \cite{raj2019probing}. Extracting these embeddings directly with a sliding window without applying VAD introduces the silence variability into the speaker embedding sequence, complicating the subsequent speaker modeling within the diarization system. Hence, to train the three proposed methods, we use the oracle VAD segmentation to the speaker embeddings beforehand, replacing the embeddings that correspond to silent segments by a zero vector $\overrightarrow{0}$ of the same dimension. We also examined the scenario where this oracle VAD segmentation was not used during training to assess whether the model 
could learn to discard the silence segments without needing the information of the VAD. 

Additionally, we evaluated the models in CallHome without using VAD, using the oracle VAD segmentation, and with an external VAD. As external VAD, we used Kaldi’s \cite{povey2011kaldi} energy-based VAD.

\section{Experimental setup}
\label{sec:setup}


\subsection{Data}

For training the models we generated the sim2spk (simulated conversations with 2 speakers) dataset, using the generation algorithm outlined in \cite{landini22_interspeech}, choosing an speech overlap ratio of 34.4\%. Specifically, sim2spk was constructed from recordings extracted from Switchboard-2 (Phases I, II, and III), Switchboard Cellular (Part1 and Part2) \cite{225858}, and NIST Speaker Recognition Evaluation (2004, 2005, 2006, and 2008) corpora \cite{przybocki04_odyssey, 912681, 4013537, martin2009nist}, all sampled at 8 kHz. As in \cite{9003959}, each utterance was augmented with background noise from the MUSAN dataset \cite{Snyder2015}, as well as having a 50\% probability of being convoluted with a randomly chosen simulated room impulse response from the RIR\footnote{https://www.openslr.org/28/} dataset \cite{rirs}. %
To adapt the models trained on simulated data to real conversations, we reserved the ``Adapt" subset of the 2-speaker CallHome CH1-2spk telephone conversation dataset \cite{martin2000nist}. 
Finally, for evaluation purposes, two distinct datasets were created: a simulated one, generated with the same process as described before; and the  remaining "Test" CH2-2spk subset of the 2-speakers CallHome dataset.

The details of the different datasets for training, testing and adaptation are shown in Table \ref{table0}.
\begin{table}[h]
\centering
\begin{tabular}{lcccc}
\toprule
\multirow{2}{*}{Datasets} & \multicolumn{2}{c}{Sim2spk} & \multicolumn{2}{c}{CallHome} \\ 
                          & Train         & Test        & Adapt         & Test         \\ 
\midrule
Total audio (h)           & 2480          & 12.4        & 3.19          & 2.97          \\ 
\midrule
Average overlap (\%)      & 34.4          & 34.4        & 14            & 13.1         \\ 
\bottomrule
\end{tabular}
\caption{Datasets used for training, adaptation and evaluation.}
\label{table0}
\end{table}

\subsection{Experimental settings}

The acoustic features extracted are 23-dimensional log-Mel-Filterbanks with a frame length of 25\,ms and a frame shift of 10\,ms. Each feature vector was concatenated with context data from the previous seven and subsequent seven frames. Concatenated features are subsampled by a factor of ten. Consequently, a $(23\times15)$-dimensional input vector is fed into the neural network every 100\,ms.

As for the speaker embeddings, 512-dimensional ECAPA-TDNN xvectors with different window sizes of 1 second, 2 and 3 seconds, and a window hop of 100\,ms are used. Experiments with concatenated features have a dimension of 857 ($23\times15 + 512$), while in the others experiments, the input dimension is the corresponding feature dimensionality. The time resolution remains constant across all the input sequences. 

We train our models using sim2spk dataset for 100 epochs. We employed the Adam optimizer with the learning rate schedule and 200,000 warm-up steps as recommended in \cite{Fujita2019EndtoEndNS}. Then we finetune these models to the 2 speakers ``Adapt" subset of CallHome for 50 epoch to adapt the models trained with simulated data to real recordings.

\subsection{System configurations}

Table \ref{paras} shows the number of parameters for the models analyzed. For each model, we use both the 3-encoder block-based model and the original 4-encoder block, with the idea that integrating the more refined speaker embeddings with respect to the acoustic features could lead to a simpler task for the encoder. The extra parameters of the method proposed in section \ref{method_a} comes from the \textit{Emb Encoder} transformer block that refines the speaker embedding sequence and reduces its dimensionality before the EDA module.

\begin{table}[h]
\centering
\begin{tabular}{lll}
\toprule
Method                                                        & \#Enc & \#Parameters \\ \midrule
\begin{tabular}[c]{@{}l@{}}EEND-EDA\\ (baseline)\end{tabular} & 4                & 6.4M         \\ \midrule
\multirow{2}{*}{Spk. Emb. into EDA}                                 & 3                & 6.5M         \\
                                                              & 4                & 7.8M         \\ \midrule
\multirow{2}{*}{Spk. Emb. into SA-EEND}                                & 3                & 5.1M         \\
                                                              & 4                & 6.4M         \\ \midrule
\multirow{2}{*}{Concat. into SA-EEND}                                  & 3                & 5.2M         \\
                                                              & 4                & 6.5M         \\ \bottomrule
\end{tabular}
\caption{Number of parameters of the baseline and the proposed methods. \textit{\#Enc} refers to the number of transformer encoder blocks in the SA-EEND encoder.}
\label{paras}
\end{table}

\subsection{Metrics}

To assess the effectiveness of the diarization process, we employed the Diarization Error Rate (DER) as metric \cite{fiscus2007rich}. Consistent with established methodologies, a collar of 0.25 seconds was permitted at the boundary of each speech segment to account for potential time misalignments. DER, which combines False Alarm Rate (FA), Missed Detection Rate (Miss) and Speaker Error Rate (SE)
, was calculated as the sum of these components divided by the total speech duration in the reference annotation:

\[
DER = \frac{FA + Miss + SE}{Total\ Speech\ Duration}
\]

This metric provides a robust measure of system performance while considering temporal variations around speech boundaries.

\section{Results}
\label{sec:results}

Our first proposed method, described in section \ref{method_a} and referred to as ``\textit{Speaker Embeddings into EDA Module,}" investigates how speaker embeddings influence the diarization performance of the EEND-EDA model when used as input to the EDA module. Results from our experiments (see table \ref{table1})  indicate that, despite an increase in model parameters,  incorporating speaker embeddings into the EDA module does not enhance performance compared to the baseline EEND-EDA model. This suggests that the speaker embeddings may not provide more discriminative information when used in this manner within the diarization task, than the one learned by the original EEND-EDA model.

\begin{table}[h]
\centering
\setlength{\tabcolsep}{5pt} 
\renewcommand{\arraystretch}{0.95} 
\begin{tabular}{ccccc}
\toprule
                   &                    & \multicolumn{3}{c}{CH2-2spk}      \\ \midrule
                   & EEND-EDA           & 8.07  & - & -    \\ \midrule
\#Enc                & w/oracle & NoVAD & OraVAD & KaldiVAD \\ \midrule
\multirow{2}{*}{3} & \ding{55}                & 20.11 & 17.96     & 19.31    \\
                   & \checkmark                 & 19.64 & 13.83     & 15.2     \\ \addlinespace
\multirow{2}{*}{4} & \ding{55}                & 18.44 & 16.75     & 17.29    \\
                   & \checkmark                & 18.32 & \textbf{11.2}      & 14.42    \\ \bottomrule
\end{tabular}
\caption{Results in DER (\%) on CH2-2spk of the \textit{Speaker embeddings into EDA module} method, described in \ref{method_a} with different system configurations, varying the number of encoder blocks, and whether the oracle VAD was used during training. The window size for extracting the embeddings is 1s.}
\label{table1}
\end{table}

\begin{table}[h]
\centering
\setlength{\tabcolsep}{4pt} 
\renewcommand{\arraystretch}{1} 
\begin{tabular}{cccccc}
\toprule
                   &                    &          & \multicolumn{3}{c}{CH2-2spk}   \\ \midrule
                   &  EEND-EDA           &          & 8.07 & - & -    \\ \midrule
\#Enc                & Size             & w/oracle & NoVAD & OraVAD & KaldiVAD \\ \midrule
\multirow{6}{*}{3} & \multirow{2}{*}{1s} & \ding{55}       & 26.77 & 23.21  & 24.96    \\
                   &                    & \checkmark      & 23.11 & \textbf{14.32}  & 16.79    \\ \addlinespace
                   & \multirow{2}{*}{2s} & \ding{55}       & 26.2  & 19.43  & 21.68    \\
                   &                    & \checkmark     & 26.74 & 18.98  & 20.11    \\ \addlinespace
                   & \multirow{2}{*}{3s} & \ding{55}       & 31.18 & 21.72  & 26.98    \\
                   &                    & \checkmark     & 23.56 & 21.15  & 22.43    \\ \addlinespace \midrule
\multirow{6}{*}{4} & \multirow{2}{*}{1s} & \ding{55}       & 32.3  & 24.28  & 28.05    \\
                   &                    & \checkmark     & 22.91 & 17.66  & 18.24    \\ \addlinespace
                   & \multirow{2}{*}{2s} & \ding{55}       & 28.1  & 20.82  & 21.35    \\
                   &                    & \checkmark     & 25.27 & 19.32  & 21.59    \\ \addlinespace
                   & \multirow{2}{*}{3s} & \ding{55}       & 30.2  & 21.39  & 27.4     \\
                   &                    & \checkmark        & 23.7  & 21.2   & 22.6     \\ \bottomrule
\end{tabular}
\caption{Results in DER (\%) on CH2-2spk for \textit{Speaker embeddings into SA-EEND encoder} method, described in \ref{method_b} with different system configurations, varying the number of the transformer encoder blocks, the window size for extracting the embeddings, and whether the oracle VAD was used during training.}
\label{table2}
\end{table}

\begin{table*}[h]
\centering
\setlength{\tabcolsep}{4pt} 
\renewcommand{\arraystretch}{0.95} 
\begin{tabular}{cccccccc}
\toprule
                   &                    &                     & \multicolumn{5}{c}{CH2-2spk}                                        \\ \midrule
                   &                    &                     & \multicolumn{2}{c}{w/o adapt} & \multicolumn{3}{c}{w/adapt}          \\ \hline
                   & EEND-EDA (Baseline)           &                     & 10.09      & -            & 8.07 & - & -        \\ \midrule
\#Enc  & Size            & w/oracle & NoVAD     & OracleVAD         & NoVAD & OracleVAD     & kaldiVAD     \\ \midrule
\multirow{6}{*}{3} & \multirow{2}{*}{1s} & \ding{55}                   &     19.23      &       14.42            &   12.87    &    8.4           &      8.79        \\
                   &                    & \checkmark                   &    16.99       &       10.39            &   12.13    &   \textbf{7.43}            &   8.07           \\ \addlinespace
                   & \multirow{2}{*}{2s} & \ding{55}                   &     17.33      &   13.81                &  12.32     &  8.64             &   8.8           \\
                   &                    & \checkmark                   &     15.34      &         \textbf{9.95}          &   13.65    &     \textbf{7.66}   & \textbf{8.03}             \\ \addlinespace
                   & \multirow{2}{*}{3s} & \ding{55}                   & 17.16     & 16.87             & 9.68  & 9.01          & 9.07         \\
                   &                    & \checkmark                   & 19.41     & \textbf{9.01}     & 13.71 & \textbf{7.62} & 8.12         \\ \addlinespace \midrule
\multirow{6}{*}{4} & \multirow{2}{*}{1s} & \ding{55}                   & 21.7      & 13.43             & 12.55 & 8.34          & 8.53         \\
                   &                    & \checkmark                   & 15.74     & \textbf{9.86}     & 13.05 & \textbf{7.2}  & \textbf{7.6} \\ \addlinespace
                   & \multirow{2}{*}{2s} & \ding{55}                   &    16.4       &                 13.93  &    11.74   & 8.45              &  8.49            \\
                   &                    & \checkmark                   &      18.98     &                 10.67  &  12.31     &        \textbf{7.4}       & \textbf{7.84}              \\ \addlinespace
                   & \multirow{2}{*}{3s} & \ding{55}                   & 15.7      & 14.9              & 9.12  & 8.8           & 9.09         \\
                   &                    & \checkmark                   & 20.44     & 11.95             & 13.76 & \textbf{7.54} & 8.56         \\ \bottomrule
\end{tabular}
\caption{Results in DER (\%) on CH2-2spk for different system configurations of the \textit{Concatenation of speaker embeddings and MFbank into SA-EEND encoder} described in \ref{method_c}, varying the number of encoder blocks, the window size used to extract the embeddings, and whether the oracle VAD was used during the model training.}
\label{table3}
\end{table*}

Our second proposed method, explained in section \ref{method_b} and referred to as ``\textit{Speaker Embeddings into SA-EEND Encoder,}" evaluates the impact of utilizing only these speaker embeddings as input to the SA-EEND encoder, and therefore as the only information provided to the whole EEND-EDA model. Table \ref{table2} displays our findings, revealing that a model configuration with 3 encoders and a 1-second window size for speaker embedding extraction yields the best results. Employing this setup during EEND-EDA training, we achieve a DER of 14.32\% when using during evaluation the oracle VAD segmentation and 16.79\% when applying an external VAD. While these results do not surpass the baseline, this approach demonstrates the significant potential of speaker embeddings as the sole input for end-to-end diarization.

\begin{figure}[h]

\includegraphics[width=\columnwidth]{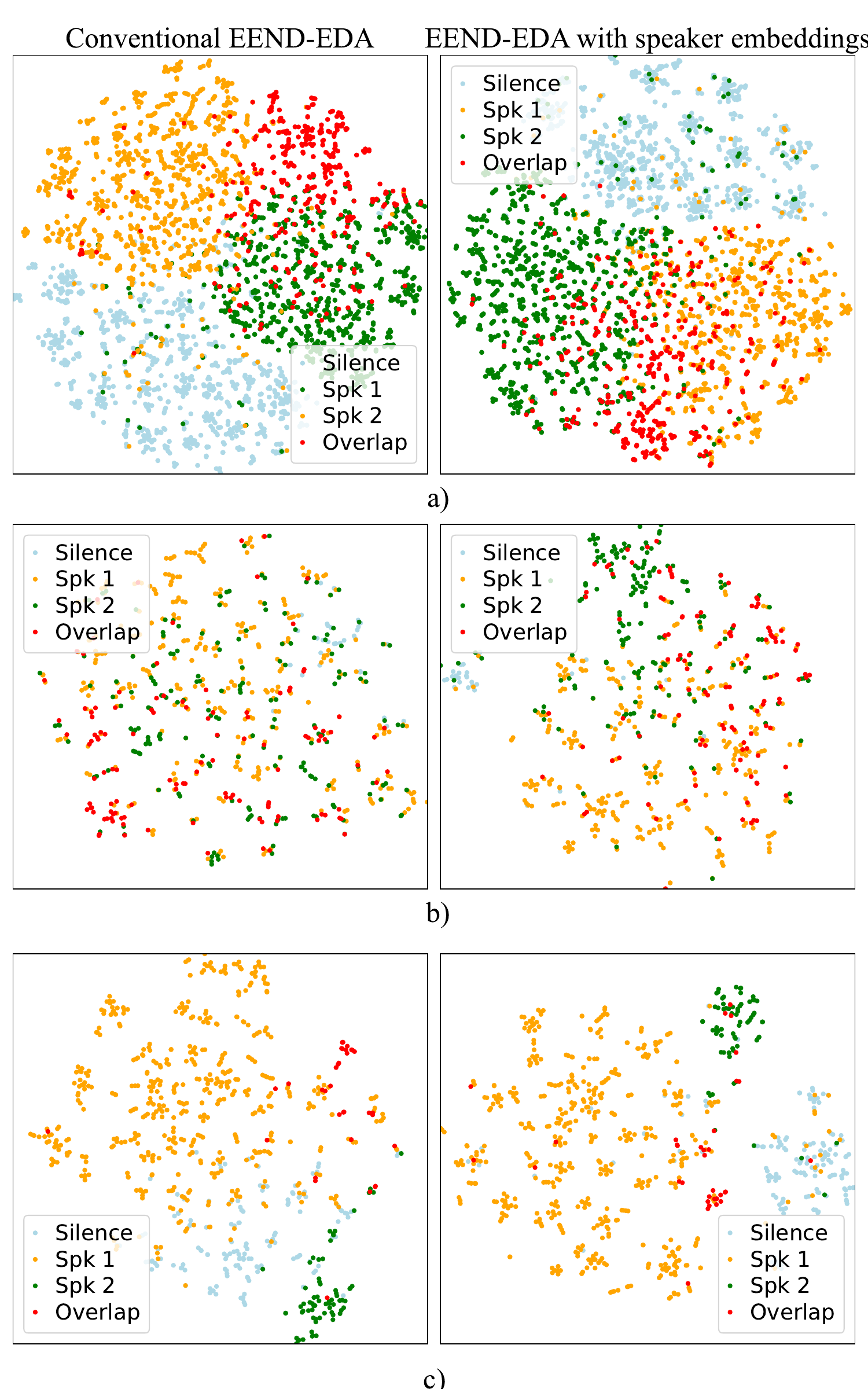}
\caption{{\it TSNE Visualization of embeddings for conventional SA-EEND-EDA and our best approach. a) Audio from simulated data, b) iafi recording, c) ialq recording.}}
\label{tres}
\end{figure}

\begin{table}[]
\centering
\begin{tabular}{cccc}
\toprule
adapt & Method & VAD & DER \\ \midrule
\multirow{5}{*}{\ding{55}} & Baseline EEND-EDA & - & 10.09 \\ \addlinespace
& \parbox{3cm}{\centering Concat EEND-EDA\\ (3 enc, 3 win len)} & Oracle & \textbf{9.01} \\ \addlinespace
& \parbox{3cm}{\centering Concat EEND-EDA\\ (4 enc, 1 win len)} & Oracle & \textbf{9.86} \\ \midrule
\multirow{5}{*}{\checkmark} & Baseline EEND-EDA & - & 8.07 \\  \addlinespace
& \parbox{3cm}{\centering Concat EEND-EDA\\ (4 enc, 1 win len)} & Oracle & \textbf{7.2} \\ \addlinespace
& \parbox{3cm}{\centering Concat EEND-EDA\\ (4 enc, 1 win len)} & Kaldi & \textbf{7.6} \\ \bottomrule
\end{tabular}
\caption{Results in DER (\%) of the performing proposed systems that correspond to the \textit{Concatenation of speaker embeddings and MFbank into
SA-EEND encoder} method using different VAD strategy. The results in CallHome ``test" set are shown before and after finetuning to the CallHome ``Adapt" subset.}
\label{fin}
\end{table}

\subsection{Results of the concatenation of speaker embeddings with acoustic features}

In Table \ref{table3}, the results obtained by the models using the proposed method described in section \ref{method_c} and so-called ``\textit{Concatenation of speaker embeddings and MFbank into SA-EEND 
encoder,}" are presented.

The performance without finetuning the models to the CallHome ``Adapt" subset is very similar to the baseline, although experiments trained and evaluated with oracle VAD segmentation, using 3 encoder blocks and a 3-second extraction window, as well as with 4 encoder blocks and a 1-second extraction window, achieve a DER of 9.01\% and 9.86\%. This represents a relative improvement over the baseline (10.09\% DER in this condition) of 10.7\% and 2.27\%, respectively.

When finetuning the models to real data using the CallHome ``Adapt" subset, and evaluating using oracle VAD segmentation, their performance is significantly improved, surpassing the baseline. Specifically, the best-performing model uses 4 encoder transformer blocks and extracts embeddings with a 1-second sized window. It achieves a DER of 7.2\%, while the baseline EEND-EDA has a performance of 8.07\% after adaptation to real conversations. This represents a relative improvement of 10.78\%. For a more realistic perspective, without oracle VAD segmentation during inference and using instead the external energy-based VAD, this model achieves a DER of 7.6\%. The relative improvement in this case reduces to 5.82\%, highlighting the importance of a reliable VAD for leveraging proper information from these embeddings in the diarization task.

Experiments where oracle VAD segmentation was not used during training, were not as satisfactory due to the silence variability leaking into the speaker embeddings. In this case, the experiment yielding the best DER uses 4 encoder layers and a 1-second window (the same configuration of the best model when trained with oracle VAD segmentation). The results obtained evaluating the model without oracle VAD is 12.55\%, whereas with oracle VAD segmentation, it is 8.34\%. It should be noted that in all experiments, whether trained with oracle VAD segmentation or not, an improvement is observed when using oracle VAD or an external VAD at evaluation time. This indicates that the EEND-EDA model's ability to discern between speech and non-speech segments is quite limited and that using external models to assist in this  task improves the overall performance. Additionally, our experiments confirm the hypothesis that using smaller window sizes for speaker embedding extraction, while potentially reducing the quality of individual speaker representations, it positively improves the overall speaker diarization performance.

Finally, the comparison between the two best-performing systems, both adapted and non-adapted, with their respective baselines is shown in table \ref{fin}.

\subsection{Visual Analysis}

In Figure \ref{tres}, the output vectors $e$ from the last SA-EEND transformer encoder block of the baseline are shown on the left, and those from the best performing of the proposed methods are shown on the right. The first row displays a randomly chosen audio file from the simulated data development set. The second row shows the \textit{iafi} audio file of the CallHome dataset, which is the most difficult audio to diarize for our model with a DER of 31.91\%, having the baseline a 48.52\% DER for this audio. The last row shows the \textit{ialq} audio file with the best DER result using our model (0\%), while the baseline provides a 1.64\% DER. In the last two cases, it can be observed that, in the proposed model, thanks to the use of an oracle VAD during training, the embeddings corresponding to silence are grouped into a more distinct cluster than in the baseline. Moreover, the distance between clusters appears to be greater, and the intra-cluster distance, for the overlap class, seems to be smaller.


\section{Conclusions}
\label{sec:conc}
In this study, we explore the integration of speaker embeddings with acoustic features for the diarization of two speakers in telephonic data within end-to-end neural diarization approaches. Our findings indicate that incorporating speaker information embeddings into the well-known EEND-EDA model leads to substantial improvements in diarization performance, in particular when using as input a concatenation of speaker embeddings with acoustic feature. Our results also showcase the importance of correctly handling silence segments and optimizing window sizes for extracting speaker information for the diarization task. The use of speaker information enhances the model’s ability to discriminate between speakers, while keeping the overlap handling capabilities of end-to-end neural systems, leading to further improvements in speaker diarization performance with respect to the EEND-EDA baseline.




\section*{Acknowledgments}

\sloppy
This work was supported by FPI RTI2018-098091-B-I00, MCIU/AEI/10.13039/501100011033/FEDER, UE and PID2021-125943OB-I00, MCIN/AEI/10.13039/501100011033/FEDER, UE from the Spanish Ministerio de Ciencia e Innovacion, Agencia y del Fondo Europeo de Desarrollo Regional.

\bibliographystyle{IEEEbib}
\bibliography{Odyssey2024_LatexTemplate}

\begin{thebibliography}{10}

\bibitem{Park2022}
Tae~Jin Park, Naoyuki Kanda, Dimitrios Dimitriadis, Kyu~J. Han, Shinji
  Watanabe, and Shrikanth Narayanan,
\newblock ``A review of speaker diarization: Recent advances with deep
  learning,''
\newblock {\em Computer Speech and Language}, vol. 72, 3 2022.

\bibitem{8461921}
Shuo-Yiin Chang, Bo~Li, Gabor Simko, Tara~N. Sainath, Anshuman Tripathi, Aäron
  van~den Oord, and Oriol Vinyals,
\newblock ``Temporal modeling using dilated convolution and gating for
  voice-activity-detection,''
\newblock in {\em 2018 IEEE International Conference on Acoustics, Speech and
  Signal Processing (ICASSP)}, 2018, pp. 5549--5553.

\bibitem{Ibrahim2018}
Noor~Salwani Ibrahim and Dzati~Athiar Ramli,
\newblock ``I-vector extraction for speaker recognition based on dimensionality
  reduction,''
\newblock {\em Procedia Computer Science}, vol. 126, pp. 1534--1540, 2018.

\bibitem{7078610}
Gregory Sell and Daniel Garcia-Romero,
\newblock ``Speaker diarization with plda i-vector scoring and unsupervised
  calibration,''
\newblock in {\em 2014 IEEE Spoken Language Technology Workshop (SLT)}, 2014,
  pp. 413--417.

\bibitem{10.1109}
Li~Wan, Quan Wang, Alan Papir, and Ignacio~Lopez Moreno,
\newblock ``Generalized end-to-end loss for speaker verification,''
\newblock in {\em 2018 IEEE International Conference on Acoustics, Speech and
  Signal Processing (ICASSP)}. 2018, p. 4879–4883, IEEE Press.

\bibitem{7953094}
Daniel Garcia-Romero, David Snyder, Gregory Sell, Daniel Povey, and Alan
  McCree,
\newblock ``Speaker diarization using deep neural network embeddings,''
\newblock in {\em 2017 IEEE International Conference on Acoustics, Speech and
  Signal Processing (ICASSP)}, 2017, pp. 4930--4934.

\bibitem{Dawalatabad2021ECAPATDNNEF}
Nauman Dawalatabad, Mirco Ravanelli, Franccois Grondin, Jenthe Thienpondt,
  Brecht Desplanques, and Hwidong Na,
\newblock ``Ecapa-tdnn embeddings for speaker diarization,''
\newblock in {\em Interspeech}, 2021.

\bibitem{Raj2019ProbingTI}
Desh Raj, David Snyder, Daniel Povey, and Sanjeev Khudanpur,
\newblock ``Probing the information encoded in x-vectors,''
\newblock {\em 2019 IEEE Automatic Speech Recognition and Understanding
  Workshop (ASRU)}, pp. 726--733, 2019.

\bibitem{dimitriadis17_interspeech}
Dimitrios Dimitriadis and Petr Fousek,
\newblock ``{Developing On-Line Speaker Diarization System},''
\newblock in {\em Proc. Interspeech 2017}, 2017, pp. 2739--2743.

\bibitem{8462628}
Quan Wang, Carlton Downey, Li~Wan, Philip~Andrew Mansfield, and Ignacio~Lopz
  Moreno,
\newblock ``Speaker diarization with lstm,''
\newblock in {\em 2018 IEEE International Conference on Acoustics, Speech and
  Signal Processing (ICASSP)}, 2018, pp. 5239--5243.

\bibitem{landini2022bayesian}
Federico Landini, J{\'a}n Profant, Mireia Diez, and Luk{\'a}{\v{s}} Burget,
\newblock ``Bayesian hmm clustering of x-vector sequences (vbx) in speaker
  diarization: theory, implementation and analysis on standard tasks,''
\newblock {\em Computer Speech \& Language}, vol. 71, pp. 101254, 2022.

\bibitem{9747450}
Youngki Kwon, Hee-Soo Heo, Jee-Weon Jung, You~Jin Kim, Bong-Jin Lee, and
  Joon~Son Chung,
\newblock ``Multi-scale speaker embedding-based graph attention networks for
  speaker diarisation,''
\newblock in {\em ICASSP 2022 - 2022 IEEE International Conference on
  Acoustics, Speech and Signal Processing (ICASSP)}, 2022, pp. 8367--8371.

\bibitem{Fujita2020EndtoEndND}
Yusuke Fujita, Shinji Watanabe, Shota Horiguchi, Yawen Xue, and Kenji
  Nagamatsu,
\newblock ``End-to-end neural diarization: Reformulating speaker diarization as
  simple multi-label classification,''
\newblock {\em ArXiv}, vol. abs/2003.02966, 2020.

\bibitem{SERAFINI2023101534}
Luca Serafini, Samuele Cornell, Giovanni Morrone, Enrico Zovato, Alessio
  Brutti, and Stefano Squartini,
\newblock ``An experimental review of speaker diarization methods with
  application to two-speaker conversational telephone speech recordings,''
\newblock {\em Computer Speech \& Language}, vol. 82, pp. 101534, 2023.

\bibitem{Fujita2019EndtoEndNS}
Yusuke Fujita, Naoyuki Kanda, Shota Horiguchi, Kenji Nagamatsu, and Shinji
  Watanabe,
\newblock ``End-to-end neural speaker diarization with permutation-free
  objectives,''
\newblock in {\em Interspeech}, 2019.

\bibitem{9003959}
Yusuke Fujita, Naoyuki Kanda, Shota Horiguchi, Yawen Xue, Kenji Nagamatsu, and
  Shinji Watanabe,
\newblock ``End-to-end neural speaker diarization with self-attention,''
\newblock in {\em 2019 IEEE Automatic Speech Recognition and Understanding
  Workshop (ASRU)}, 2019, pp. 296--303.

\bibitem{Horiguchi2020EndtoEndSD}
Shota Horiguchi, Yusuke Fujita, Shinji Watanabe, Yawen Xue, and Kenji
  Nagamatsu,
\newblock ``End-to-end speaker diarization for an unknown number of speakers
  with encoder-decoder based attractors,''
\newblock in {\em Interspeech}, 2020.

\bibitem{eendeda}
Shota Horiguchi, Yusuke Fujita, Shinji Watanabe, Yawen Xue, and Paola Garcia,
\newblock ``Encoder-decoder based attractors for end-to-end neural
  diarization,''
\newblock {\em IEEE/ACM Transactions on Audio, Speech, and Language
  Processing}, vol. 30, pp. 1--1, 01 2022.

\bibitem{9414333}
Keisuke Kinoshita, Marc Delcroix, and Naohiro Tawara,
\newblock ``Integrating end-to-end neural and clustering-based diarization:
  Getting the best of both worlds,''
\newblock in {\em ICASSP 2021 - 2021 IEEE International Conference on
  Acoustics, Speech and Signal Processing (ICASSP)}, 2021, pp. 7198--7202.

\bibitem{10095589}
Ye-Rin Jeoung, Joon-Young Yang, Jeong-Hwan Choi, and Joon-Hyuk Chang,
\newblock ``Improving transformer-based end-to-end speaker diarization by
  assigning auxiliary losses to attention heads,''
\newblock in {\em ICASSP 2023 - 2023 IEEE International Conference on
  Acoustics, Speech and Signal Processing (ICASSP)}, 2023, pp. 1--5.

\bibitem{Chen2023}
Zhengyang Chen, Bing Han, Shuai Wang, and Yanmin Qian,
\newblock ``Attention-based encoder-decoder network for end-to-end neural
  speaker diarization with target speaker attractor,''
\newblock in {\em Proc. Interspeech 2023}, 08 2023, pp. 3552--3556.

\bibitem{10022924}
Soumi Maiti, Yushi Ueda, Shinji Watanabe, Chunlei Zhang, Meng Yu, Shi-Xiong
  Zhang, and Yong Xu,
\newblock ``Eend-ss: Joint end-to-end neural speaker diarization and speech
  separation for flexible number of speakers,''
\newblock in {\em 2022 IEEE Spoken Language Technology Workshop (SLT)}, 2023,
  pp. 480--487.

\bibitem{10095370}
Tobias Cord-Landwehr, Christoph Boeddeker, Cătălin Zorilă, Rama Doddipatla,
  and Reinhold Haeb-Umbach,
\newblock ``Frame-wise and overlap-robust speaker embeddings for meeting
  diarization,''
\newblock in {\em ICASSP 2023 - 2023 IEEE International Conference on
  Acoustics, Speech and Signal Processing (ICASSP)}, 2023, pp. 1--5.

\bibitem{xia2022turn}
Wei Xia, Han Lu, Quan Wang, Anshuman Tripathi, Yiling Huang, Ignacio~Lopez
  Moreno, and Hasim Sak,
\newblock ``Turn-to-diarize: Online speaker diarization constrained by
  transformer transducer speaker turn detection,''
\newblock in {\em Proc. ICASSP}, 2022, pp. 8077--8081.

\bibitem{10094955}
Guanlong Zhao, Quan Wang, Han Lu, Yiling Huang, and Ignacio~Lopez Moreno,
\newblock ``Augmenting transformer-transducer based speaker change detection
  with token-level training loss,''
\newblock in {\em ICASSP 2023 - 2023 IEEE International Conference on
  Acoustics, Speech and Signal Processing (ICASSP)}, 2023, pp. 1--5.

\bibitem{Medennikov2020}
Ivan Medennikov, Maxim Korenevsky, Tatiana Prisyach, Yuri Khokhlov, Maria
  Korenevskaya, Ivan Sorokin, Tatiana Timofeeva, Anton Mitrofanov, Andrei
  Andrusenko, Ivan Podluzhny, Aleksandr Laptev, and Aleksei Romanenko,
\newblock ``Target-speaker voice activity detection: A novel approach for
  multi-speaker diarization in a dinner party scenario,''
\newblock in {\em Interspeech}, 10 2020, pp. 274--278.

\bibitem{9747772}
Weiqing Wang and Ming Li,
\newblock ``Incorporating end-to-end framework into target-speaker voice
  activity detection,''
\newblock in {\em ICASSP 2022 - 2022 IEEE International Conference on
  Acoustics, Speech and Signal Processing (ICASSP)}, 2022, pp. 8362--8366.

\bibitem{10095185}
Dongmei Wang, Xiong Xiao, Naoyuki Kanda, Takuya Yoshioka, and Jian Wu,
\newblock ``Target speaker voice activity detection with transformers and its
  integration with end-to-end neural diarization,''
\newblock in {\em ICASSP 2023 - 2023 IEEE International Conference on
  Acoustics, Speech and Signal Processing (ICASSP)}, 2023, pp. 1--5.

\bibitem{conformer}
Yi~Liu, Eunjung Han, Chul Lee, and Andreas Stolcke,
\newblock ``End-to-end neural diarization: From transformer to conformer,''
\newblock in {\em Interspeech}, 08 2021, pp. 3081--3085.

\bibitem{7952154}
Dong Yu, Morten Kolbæk, Zheng-Hua Tan, and Jesper Jensen,
\newblock ``Permutation invariant training of deep models for
  speaker-independent multi-talker speech separation,''
\newblock in {\em 2017 IEEE International Conference on Acoustics, Speech and
  Signal Processing (ICASSP)}, 2017, pp. 241--245.

\bibitem{8953658}
Jiankang Deng, Jia Guo, Niannan Xue, and Stefanos Zafeiriou,
\newblock ``Arcface: Additive angular margin loss for deep face recognition,''
\newblock in {\em 2019 IEEE/CVF Conference on Computer Vision and Pattern
  Recognition (CVPR)}, 2019, pp. 4685--4694.

\bibitem{ioffe2006probabilistic}
Sergey Ioffe,
\newblock ``Probabilistic linear discriminant analysis,''
\newblock in {\em Computer Vision--ECCV 2006: 9th European Conference on
  Computer Vision, Graz, Austria, May 7-13, 2006, Proceedings, Part IV 9}.
  Springer, 2006, pp. 531--542.

\bibitem{nagrani2017voxceleb}
Arsha Nagrani, Joon~Son Chung, and Andrew Zisserman,
\newblock ``Voxceleb: a large-scale speaker identification dataset,''
\newblock {\em arXiv preprint arXiv:1706.08612}, 2017.

\bibitem{chung2018voxceleb2}
Joon~Son Chung, Arsha Nagrani, and Andrew Zisserman,
\newblock ``Voxceleb2: Deep speaker recognition,''
\newblock {\em arXiv preprint arXiv:1806.05622}, 2018.

\bibitem{Snyder2015}
David Snyder, Guoguo Chen, and Daniel Povey,
\newblock ``Musan: A music, speech, and noise corpus,''
\newblock 10 2015.

\bibitem{raj2019probing}
Desh Raj, David Snyder, Daniel Povey, and Sanjeev Khudanpur,
\newblock ``Probing the information encoded in x-vectors,''
\newblock in {\em 2019 IEEE Automatic Speech Recognition and Understanding
  Workshop (ASRU)}. IEEE, 2019, pp. 726--733.

\bibitem{povey2011kaldi}
Daniel Povey, Arnab Ghoshal, Gilles Boulianne, Lukas Burget, Ondrej Glembek,
  Nagendra Goel, Mirko Hannemann, Petr Motlicek, Yanmin Qian, Petr Schwarz,
  et~al.,
\newblock ``The kaldi speech recognition toolkit,''
\newblock in {\em IEEE 2011 workshop on automatic speech recognition and
  understanding}. IEEE Signal Processing Society, 2011, number CONF.

\bibitem{landini22_interspeech}
Federico Landini, Alicia Lozano-Diez, Mireia Diez, and Lukáš Burget,
\newblock ``{From Simulated Mixtures to Simulated Conversations as Training
  Data for End-to-End Neural Diarization},''
\newblock in {\em Proc. Interspeech 2022}, 2022, pp. 5095--5099.

\bibitem{225858}
J.J. Godfrey, E.C. Holliman, and J.~McDaniel,
\newblock ``Switchboard: telephone speech corpus for research and
  development,''
\newblock in {\em [Proceedings] ICASSP-92: 1992 IEEE International Conference
  on Acoustics, Speech, and Signal Processing}, 1992, vol.~1, pp. 517--520
  vol.1.

\bibitem{przybocki04_odyssey}
Mark Przybocki and Alvin~F. Martin,
\newblock ``{NIST speaker recognition evaluation chronicles},''
\newblock in {\em Proc. The Speaker and Language Recognition Workshop (Odyssey
  2004)}, 2004, pp. 15--22.

\bibitem{912681}
Omid Sadjadi, Craig Greenberg, Elliot Singer, Lisa Mason, and Douglas Reynolds,
\newblock ``Nist 2021 speaker recognition evaluation plan,'' 2021-07-12
  04:07:00 2021.

\bibitem{4013537}
Mark~A. Przybocki, Alvin~F. Martin, and Audrey~N. Le,
\newblock ``Nist speaker recognition evaluation chronicles - part 2,''
\newblock in {\em 2006 IEEE Odyssey - The Speaker and Language Recognition
  Workshop}, 2006, pp. 1--6.

\bibitem{martin2009nist}
Alvin~F Martin and Craig~S Greenberg,
\newblock ``Nist 2008 speaker recognition evaluation: Performance across
  telephone and room microphone channels,''
\newblock in {\em Tenth Annual Conference of the International Speech
  Communication Association}, 2009.

\bibitem{rirs}
Tom Ko, Vijayaditya Peddinti, Daniel Povey, Michael~L. Seltzer, and Sanjeev
  Khudanpur,
\newblock ``A study on data augmentation of reverberant speech for robust
  speech recognition,''
\newblock in {\em 2017 IEEE International Conference on Acoustics, Speech and
  Signal Processing (ICASSP)}, 2017, pp. 5220--5224.

\bibitem{martin2000nist}
Alvin Martin and Mark Przybocki,
\newblock ``The nist 1999 speaker recognition evaluation—an overview,''
\newblock {\em Digital signal processing}, vol. 10, no. 1-3, pp. 1--18, 2000.

\bibitem{fiscus2007rich}
Jonathan~G Fiscus, Jerome Ajot, and John~S Garofolo,
\newblock ``The rich transcription 2007 meeting recognition evaluation,''
\newblock in {\em International Evaluation Workshop on Rich Transcription}.
  Springer, 2007, pp. 373--389.

\end{thebibliography}

\end{document}